%% file: main.tex
\documentclass[10pt,conference]{IEEEtran}
\IEEEoverridecommandlockouts
\usepackage{cite}
\usepackage{amsmath,amssymb,amsfonts}
\usepackage{algorithmic}
\usepackage{graphicx}
\usepackage{textcomp}
\usepackage{xcolor}
\usepackage{rotating}
\usepackage{float}

\def\BibTeX{{\rm B\kern-.05em{\sc i\kern-.025em b}\kern-.08em
    T\kern-.1667em\lower.7ex\hbox{E}\kern-.125emX}}

\newcommand{\RN}[1]{%
	\textup{\uppercase\expandafter{\romannumeral#1}}%
}

\usepackage[symbol]{footmisc}

\input{preamble}

\begin{document}

\title{Shallow or Deep? An Empirical Study on Detecting
Vulnerabilities using Deep Learning}

\author{\IEEEauthorblockN{Alejandro Mazuera-Rozo\textsuperscript{\textdagger}$^{1,2}$ \thanks{\textsuperscript{\textdagger} Both authors contributed equally to this manuscript.}, Anamaria Mojica-Hanke\textsuperscript{\textdagger}$^{2}$, Mario~Linares-V\'asquez$^{2}$, Gabriele Bavota$^{1}$}
	\IEEEauthorblockA{
		\textit{$^{1}$SEART @ Software Institute, Universit\`a della Svizzera italiana, Lugano, Switzerland} \\
		\textit{$^{2}$Universidad de los Andes, Bogot\'a, Colombia}\vspace{-0.5cm}}

}

\maketitle

\begin{abstract}
Deep learning (DL) techniques are on the rise in the software engineering research community. More and more approaches have been developed on top of DL models, also due to the unprecedented amount of software-related data that can be used to train these models. One of the recent applications of DL in the software engineering domain concerns the automatic detection of software vulnerabilities. While several DL models have been developed to approach this problem, there is still limited empirical evidence concerning their actual effectiveness especially when compared with shallow machine learning techniques.
In this paper, we partially fill this gap by presenting a large-scale empirical study using three vulnerability datasets and five different source code representations (\ie the format in which the code is provided to the classifiers to assess whether it is vulnerable or not) to compare the effectiveness of two widely used DL-based models and of one shallow machine learning model in (i) classifying code functions as vulnerable or non-vulnerable (\ie binary classification), and (ii) classifying code functions based on the specific type of vulnerability they contain (or ``clean'', if no vulnerability is there). As a baseline we include in our  study the AutoML utility provided by the Google Cloud Platform. Our results show that the experimented models are still far from ensuring reliable vulnerability detection, and that a shallow learning classifier represents a competitive baseline for the newest DL-based models.

\end{abstract}


\input{secs/introduction}
\input{secs/related_work}
\input{secs/empirical_study}
\input{secs/results}
\input{secs/threats}

\input{secs/conclusion}

\vspace{0.2cm}
\section{Data Availability}
The data used in our study are publicly available \cite{replication}.

\vspace{0.2cm}
\input{ack}

\newpage

\balance
\bibliographystyle{IEEEtran}
\bibliography{bib/main}

\end{document}

%% file: preamble.tex
\usepackage{amssymb}
\usepackage{hyperref}
\usepackage[plain]{fancyref}
\usepackage{ifdraft}

\usepackage[inline]{enumitem}
\usepackage{xcolor}
\usepackage{xspace}
\usepackage[final]{listings}
\usepackage{acronym}
\usepackage{url}
\usepackage{amsmath}
\usepackage{amssymb}
\usepackage{booktabs} 
\usepackage{subfig}
\usepackage{balance}
\usepackage{dirtree}
\usepackage{multirow}

\usepackage[ruled]{algorithm2e} 

\newcommand{\subparagraph}{}
\usepackage[compact]{titlesec}
\titlespacing{\section}{0pt}{*0}{*0}
\titlespacing{\subsection}{0pt}{*0}{*0}
\titlespacing{\subsubsection}{0pt}{*0}{*0}

\usepackage{etoolbox}
\makeatletter
\patchcmd{\@makecaption}
  {\scshape}
  {}
  {}
  {}
\patchcmd{\@makecaption}
  {\\}
  {.\ }
  {}
  {}
\makeatother

\definecolor{OliveGreen}{rgb}{0,0.6,0.3}

\renewcommand{\lstlistingname}{Snippet}
\newcommand*{\fancyreflstlabelprefix}{lst}
\newcommand*{\Freflstname}{\lstlistingname}
\newcommand*{\freflstname}{\MakeLowercase{\lstlistingname}}
\Frefformat{vario}{\fancyreflstlabelprefix}%
  {\Freflstname\fancyrefdefaultspacing#1#3}
\frefformat{vario}{\fancyreflstlabelprefix}%
  {\freflstname\fancyrefdefaultspacing#1#3}
\Frefformat{plain}{\fancyreflstlabelprefix}%
  {\Freflstname\fancyrefdefaultspacing#1}
\frefformat{plain}{\fancyreflstlabelprefix}%
  {\freflstname\fancyrefdefaultspacing#1}

\renewcommand{\tablename}{Table}  
  
\newcommand*{\fancyreflnlabelprefix}{ln}
\newcommand*{\Freflnname}{Line}
\newcommand*{\freflnname}{\MakeLowercase{\Freflnname}}
\Frefformat{vario}{\fancyreflnlabelprefix}%
  {\Freflnname\fancyrefdefaultspacing#1#3}
\frefformat{vario}{\fancyreflnlabelprefix}%
  {\freflnname\fancyrefdefaultspacing#1#3}
\Frefformat{plain}{\fancyreflnlabelprefix}%
  {\Freflnname\fancyrefdefaultspacing#1}
\frefformat{plain}{\fancyreflnlabelprefix}%
  {\freflnname\fancyrefdefaultspacing#1}

\lstdefinelanguage{JavaScript}{
keywords={typeof, new, true, false, catch, function, return, null, catch, switch, var, if, in, for, while, do, else, case, break, throw, this, instanceof},
keywordstyle=\color{purple}\bfseries,
ndkeywords={},
ndkeywordstyle=\color{blue}\bfseries,
identifierstyle=\color{black},
sensitive=false,
comment=[l]{//},
morecomment=[s]{/*}{*/},
commentstyle=\color{OliveGreen}\ttfamily,
stringstyle=\color{OliveGreen}\ttfamily,
morestring=[b]',
morestring=[b]"
}
\usepackage{color}
\definecolor{gray97}{gray}{.97}
\definecolor{gray90}{gray}{.90}
\definecolor{gray75}{gray}{.75}
\definecolor{gray45}{gray}{.45}
\definecolor{codegreen}{rgb}{0,0.6,0}
\definecolor{codered}{rgb}{0.6,0,0}
\definecolor{codegray}{rgb}{0.5,0.5,0.5}
\definecolor{codepurple}{rgb}{0.58,0,0.82}
\lstdefinestyle{floating}{%
  frame=none,
  float=htb,
  captionpos=b
}

\lstdefinestyle{ctxtraits}
 {language=JavaScript,
  frame=lines,
  showstringspaces=false,
  keywordstyle=\tt\bf,
  tabsize=3,
  style=floating,
  morekeywords={Trait, cop, Context, activate, deactivate, adapt, addObjectPolicy, manager}
}

\lstnewenvironment{ctxtraits}[1][]
 {\lstset{style=ctxtraits,#1}}{}

\newcommand{\eg}{\emph{e.g.,}\xspace}
\newcommand{\ie}{\emph{i.e.,}\xspace}
\newcommand{\etal}{\emph{et al.}\xspace}

\newcommand{\secref}[1]{Section~\ref{#1}\xspace}
\newcommand{\figref}[1]{Fig.~\ref{#1}\xspace}

\newcommand{\tabref}[1]{Table~\ref{#1}\xspace}

\definecolor{author}{rgb}{.5, .5, .5}
\definecolor{comment}{rgb}{.1, .0, .9}
\definecolor{note}{rgb}{.9, .4, .0}
\definecolor{idea}{rgb}{.1, .7, .0}
\definecolor{missing}{rgb}{.9, .1, .0}
\definecolor{deleteme}{rgb}{.9, .1, .0}

\input{acronyms}

%% file: acronyms.tex
\acrodef{APK}{Android Application Package}

%% file: secs/introduction.tex
\section{Introduction}

\label{sec:intro}
DL models have been used to support software-related tasks such as bug localization \cite{DBLP:conf/iwpc/LamNNN17}, automated bug fixing \cite{Tufano:2018:EIL:3238147.3240732,Chen:2019,Mesbah:2019:DLR:3338906.3340455,DBLP:journals/corr/abs-1812-07170}, clone detection \cite{WhiteTVP16}, code search \cite{DeepCodeSearch}, and code summarization \cite{LeClair:2019:NMG:3339505.3339605,Liu:2018:NCM:3238147.3238190,inproceedingsMine}. DL models have also been used to automatically detect software vulnerabilities \cite{RUSSELL, FWU2017, Li20182, HARER2018, Li2018, ZOU2019, WANG2019, LIN2019}. While several of these studies present a thorough evaluation of the technique they propose, there is still limited empirical evidence about the advantages brought by DL models over shallow machine learning algorithms. We present a large-scale study comparing two widely used deep learning models, namely Convolutional Neural Network (CNN) and Recurrent Neural Network (RNN), with the Random Forest classifier, one of the most popular ``shallow learning'' models.
The comparison of these three learning models is conducted across three different datasets including vulnerable and non-vulnerable C/C++ functions. Two of the employed datasets have been previously used in the literature to evaluate the effectiveness of DL-based vulnerability detection techniques \cite{JULIET,RUSSELL}. The first one \cite{JULIET} includes synthetic code examples (\ie the vulnerabilities have been manually injected) of 118 types of software vulnerabilities. The second \cite{RUSSELL} includes functions from the Debian Linux distribution and GitHub public repositories that have been labeled as vulnerable or non-vulnerable by using static analysis tools (\ie Clang \cite{CLANG}, Cppcheck \cite{CPPCHECK}, and Flawfinder \cite{FLAWF}). 

Since the first dataset \cite{JULIET} is composed of artificial instances, and the second one \cite{RUSSELL}, while including real code, is \textbf{highly unbalanced} ($>$90\% of its functions are non-vulnerable), we built a third \textbf{balanced dataset} composed of real instances mined from GitHub. Our work builds on top of the seminal paper by Russell \etal \cite{RUSSELL}, by testing/investigating DL-based vulnerability detection techniques on an additional dataset that  is balanced and composed of real code vulnerability instances,  making it more suitable for experiments involving learning techniques. 
Overall, our empirical study includes a total of $\sim$ 1.8M C/C++ functions.

Besides using different datasets, we also experiment the effectiveness of different code representations (\ie the format in which the source code is provided as input to the learning models). We consider five different representations inspired by the work of Tufano \etal \cite{Tufano:2018:EIL:3238147.3240732} and using different levels of source code abstraction. Finally, we executed the experiments also with the AutoML utility provided by Google Cloud Platform. The study of different code representations as well as the usage of AutoML as a baseline for DL methods represent, besides the novel dataset we contribute to the community, additional points of novelty for our work as compared to the most related studies \cite{RUSSELL, Li2018}.

Our work does not aim to replicate or verify the reproducibility of previous approaches presented in the literature for DL-based vulnerability detection. Indeed, many factors come into play when implementing, training, and customizing the developed models. Our goal is to empirically evaluate our own implementations of the DL and shallow learning models using different datasets and code representations, to assess the advantages (if any) brought by the DL-models over the shallow one. The material used in our study is publicly available \cite{replication}. 

The achieved results show that (i) on the dataset featuring synthetic examples of vulnerabilities \cite{JULIET}, all experimented models behave almost as perfect predictors, questioning the usefulness of this dataset for assessing vulnerability predictors; (ii) on the other datasets, as expected, the models struggle to achieve very high performance, showing that important margins for improvement are possible; and (iii) the shallow model represents a very competitive approach for the DL-based models we tested.

%% file: secs/related_work.tex

\section{Related Work}

\label{sec:related}
There are several studies describing and analyzing vulnerabilities and their impact in software projects \cite{LinaresVasquez2017, MazueraRozo2019, Barabanov2018, Rafique2015, Huang2010, Islam2017, Homaei2017, Nappa2015, Papp2015, KOTZIAS}. Given the goal of our work, we focus our discussion on approaches to detect software vulnerabilities in code. The latter can be roughly classified into three groups: (i) approaches using static and dynamic code analyses with detection rules to identify vulnerabilities; (ii) program analysis combined with shallow learning techniques (\ie traditional machine learning classifiers/predictors); and (iii) more recent approaches for learning semantic and syntactic features of vulnerable code using deep learning. Given the very extensive literature in this field, we focus our discussion on a limited number of representative examples for each of these three categories.

Approaches based on static and dynamic code analysis are the most diffused. Those relying on static analysis examine the code without executing the program. A representative example of this family of techniques is the work by Pengfei \etal \cite{Pengfei2017} that uses pattern-based matching to detect double fetch vulnerabilities in the Linux kernel;  other examples of available tools for static analysis are Clang \cite{CLANG}, Flawfinder \cite{FLAWF}, Checkmarx \cite{Checkmarx} and Cppcheck \cite{CPPCHECK}. Dynamic analysis-based detectors assess a program by injecting data in real-time or simulating conditions that could trigger states leading to vulnerabilities (see \eg AddressSanitizer \cite{ASAN}, SANTE \cite{Chebaro2013}, DR. CHECKER \cite{Aravind2017}, OPIA \cite{OPIA}, and DroidForensics \cite{Yuan2017}). Moreover, hybrid software techniques have also been proposed: Concolic testing approaches use dynamic symbolic execution to reach a trade off between the costs and benefits of dynamic and static analysis. Tools such as FUZZBUSTER \cite{MUSLINER2012}, VUzzer \cite{RAWAT2017} and QSYM \cite{YUN2018} are examples of this trend.

Concerning techniques using shallow learning, a representative example is the work by Hovsepyan \etal \cite{Hovsepyan}, in which the authors use a support vector machine (SVM) to train a vulnerabilities detector; Hovsepyan \etal use a tokenized representation of Java source code to predict whether a Java file is vulnerable or not. Another representative approach is SOFIA \cite{Ceccato2016}, which is a programming-language and source-code independent Security Oracle; the approach is unsupervised and uses clustering to group SQL statements and detect anomalies possibly representing security issues. Lastly, Perl \etal \cite{Perl2015} presented VCCFinder, an approach to find potentially dangerous code in software repositories; VCCFinder works on code snippets to detect suspicious commits by using SVM. 

\begin{table*}[h]
\caption{Previous work using deep learning for vulnerability detection.}
\label{tab:related-work}
\resizebox{18cm}{!} {
\begin{tabular}{llllllll}
\hline
\multicolumn{1}{c}{\textbf{Study}} & \multicolumn{1}{c}{\textbf{Data origin}} & \multicolumn{1}{c}{\textbf{Lang.}} & \multicolumn{1}{c}{\textbf{Assessed Artefact}} & \multicolumn{1}{c}{\textbf{Representation}} & \multicolumn{1}{c}{\textbf{Classifiers}} & \multicolumn{1}{c}{\textbf{Classification}} & \multicolumn{1}{c}{\textbf{Vulnerabilities}} \\ \hline

\begin{tabular}[c]{@{}l@{}}Russell \etal \cite{RUSSELL}\end{tabular} & \begin{tabular}[c]{@{}l@{}}SATE IV, Github \\ \& Debian open \\ repositories\end{tabular} & C/C++ & \begin{tabular}[c]{@{}l@{}}${\sim}$1.3M  Functions\end{tabular} & \begin{tabular}[c]{@{}l@{}}Relevant meaning of \\ critical tokens  (\eg keywords, \\ operators and  separators)\end{tabular} & \begin{tabular}[c]{@{}l@{}}BoW+RF, RNN, \\ CNN, RNN+RF, \\ CNN+RF\end{tabular} & \begin{tabular}[c]{@{}l@{}}Multiple Binary \\ classification\end{tabular} & \begin{tabular}[c]{@{}l@{}}CWE-120, CWE-119, \\ CWE-476, CWE-469,  \\ CWE-Others\end{tabular} \\ \hline

\begin{tabular}[c]{@{}l@{}}Wu \etal \cite{FWU2017}\end{tabular} & \begin{tabular}[c]{@{}l@{}}Binary programs \\ in "/src/bin" and\\  "/usr/sbin/"\end{tabular} & C & \begin{tabular}[c]{@{}l@{}}${\sim}$10K Sequences\\  of function calls\end{tabular} & \begin{tabular}[c]{@{}l@{}}Events regarding sequential calls\\  within the program augmented \\ with its arguments\end{tabular} & \begin{tabular}[c]{@{}l@{}}CNN, LSTM, \\ CNN-LSTM, MLP\end{tabular} & \begin{tabular}[c]{@{}l@{}}Binary \\ classification\end{tabular} & None \\ \hline

\begin{tabular}[c]{@{}l@{}}Li \etal \cite{Li20182}\end{tabular} & \begin{tabular}[c]{@{}l@{}}NVD open source \\ software programs \\ \& SARD\end{tabular} & C/C++ & ${\sim}$420k SeVCs & \begin{tabular}[c]{@{}l@{}}Sliced program representations \\ that can accommodate syntax and \\ semantic information pertinent\\  to vulnerabilities\end{tabular} & \begin{tabular}[c]{@{}l@{}}CNN, DBN, LSTM, \\ GRU, BLSTM, BGRU\end{tabular} & \begin{tabular}[c]{@{}l@{}}multi-class \\ classification\\  and Binary \\ classification\end{tabular} & \begin{tabular}[c]{@{}l@{}}4 Kinds ( \ie issues related \\ to library/API function calls, \\ arrays, pointers and improper \\ arithmetic expressions)\end{tabular} \\ \hline

\begin{tabular}[c]{@{}l@{}}Harer \etal \cite{HARER2018}\end{tabular} & \begin{tabular}[c]{@{}l@{}}Packages distributed\\  with the Debian Linux \\ distribution and functions \\ pulled from public Git \\ repositories on Github\end{tabular} & C/C++ & ${\sim}$900k Functions & \begin{tabular}[c]{@{}l@{}}Parsed code and categorized \\ elements into different bins \\ (\eg string literals, numbers, \\ operators)\end{tabular} & \begin{tabular}[c]{@{}l@{}}Word2vec+ CNN, \\ CNN+ET, BOW+ET\end{tabular} & \begin{tabular}[c]{@{}l@{}}Binary \\ classification\end{tabular} & None \\ \hline

\begin{tabular}[c]{@{}l@{}}Li \etal \cite{Li2018}\end{tabular} & \begin{tabular}[c]{@{}l@{}}NVD open source \\ software programs \\ \& SARD\end{tabular} & C/C++ & \begin{tabular}[c]{@{}l@{}}${\sim}$61k Code gadgets\end{tabular} & \begin{tabular}[c]{@{}l@{}}Program slices represented  as \\ lines of code that are semantically \\ related to each other\end{tabular} & BLSTM & \begin{tabular}[c]{@{}l@{}}Multiple Binary \\ classification\end{tabular} & CWE-119, CWE-399 \\ \hline

\begin{tabular}[c]{@{}l@{}}Zou \etal \cite{ZOU2019}\end{tabular} & \begin{tabular}[c]{@{}l@{}}NVD open source\\  software programs \\ \& SARD\end{tabular} & C/C++ & \begin{tabular}[c]{@{}l@{}}Code attention \& Code \\ gadgets extracted from  \\ ${\sim}$33k Programs\end{tabular} & \begin{tabular}[c]{@{}l@{}}Multiple program semantically \\ related statements in a piece of \\ code being discrete aspects of \\ information\end{tabular} & Enhanced BLSTM & \begin{tabular}[c]{@{}l@{}}Multi-class \\ classification\end{tabular} & \begin{tabular}[c]{@{}l@{}}40 types of vulnerabilities \\ (\eg CWE-119)\end{tabular} \\ \hline

\begin{tabular}[c]{@{}l@{}}Wang \etal  \cite{WANG2019}\end{tabular} & \begin{tabular}[c]{@{}l@{}}SARD, GitHub \\ \& Exploit-DB\end{tabular} & C/C++ & \begin{tabular}[c]{@{}l@{}}Execution paths retrieved \\ from  ${\sim}$56k binary \\ programs\end{tabular} & \begin{tabular}[c]{@{}l@{}}Vector representation retaining \\ original semantic information \\ of the execution path\end{tabular} & LSTM & \begin{tabular}[c]{@{}l@{}}Binary \\ classification\end{tabular} & None \\ \hline

\begin{tabular}[c]{@{}l@{}}Lin \etal \cite{LIN2019}\end{tabular} & \begin{tabular}[c]{@{}l@{}}SARD \& Real-world \\ Open source projects \\ (\eg LibTIFF, LibPNG)\end{tabular} & C/C++ & ${\sim}$169k Functions & \begin{tabular}[c]{@{}l@{}}Vectors representing sequences \\ computed from the source \\ code and ASTs\end{tabular} & RNN (Bi-LSTM) & \begin{tabular}[c]{@{}l@{}}Binary \\ classification\end{tabular} & None \\ \hline
\end{tabular}
}
\end{table*}

The deep learning (DL) approaches also use classification as a way to categorize code as vulnerable or not, similarly to what shallow learning techniques do. However, in the case of DL, compositional representations are automatically learnt.  When using DL, semantic and syntactic features can be automatically extracted from the source code into vectors representing the code and can be fed into a neural network. Such a flexibility has led to the application of DL to many software engineering tasks (see \secref{sec:intro} for a list of works applying DL in the software engineering domain). One of these tasks is the detection of security vulnerabilities. Some representative works in this area are discussed in the following. 

The interested reader can refer to the work by Lin \etal \cite{Lin2020} for a complete overview of the relevant literature in the field. Most of the DL-based approaches  for detecting vulnerabilities have used Convolutional Neural Network (CNN) \cite{RUSSELL, FWU2017, Li20182, HARER2018}, Recurrent Neural Network (RNN) \cite{RUSSELL, Li2018, Li20182, LIN2019} or improved variants of these models \cite{FWU2017, Li2018, Li20182, ZOU2019, WANG2019}. \tabref{tab:related-work} summarizes those works. 

For instance, Russell \etal \cite{RUSSELL} use CNN with custom lexed representations of C/C++ source code, working with a dataset of over 1.2 million curated functions. Li \etal \cite{Li2018} propose VulDeePecker, a deep learning-based system for vulnerability detection using BLSTM; it uses their own fine-grained representation of C/C++ code called \textit{code gadget}, which is a set of lines of code that are semantically related. VulDeePecker was validated with two types of vulnerabilities, buffer error vulnerabilities (CWE-119) and resource management error vulnerabilities (CWE-399). Li \etal \cite{Li20182} have presented a new approach outperforming VulDeePecker, by implementing a Bidirectional Gated Recurrent Unit (BGRU) model which overcomes the results of other state-of-the-art detectors and uses a new representation of C/C++ source code that considers syntactic and semantic information; this model uses 126 types of vulnerabilities aggregated in 4 categories: issues related to library/API function calls, arrays, pointers and improper arithmetic expressions. Zou \etal \cite{ZOU2019} presented µVulDeePecker, a deep learning-based system for multi-class vulnerability detection, considering a total of 40 different types of software vulnerabilities.

In our work, we experiment with implementations of an RNN-based and a CNN-based model as representative of the work done in the field of DL-based vulnerability detection, and a Random Forest classifier as representative of shallow learning techniques.

%% file: secs/empirical_study.tex
\section{Empirical Study} \label{sec:design}

\newcommand\rqone{What is the  effectiveness of different  combinations of classifiers and code representations to identify functions affected by software vulnerabilities?}

The \emph{goal} of this study is to assess the effectiveness of deep/shallow learning techniques for detecting software vulnerabilities at function-level granularity when using different models and source code abstractions. We conduct experiments with three different models (two deep and one shallow). In particular, we experiment with: (i) Random Forest (RF), (ii) a Convolutional Neural Network (CNN), and (iii) a Recurrent Neural Network (RNN), with the first being representative of shallow classifiers and the last two of deep learning models. We chose RF due to its popularity in the software engineering domain (see \eg \cite{Selby1988,Porter1990,Briand1992}). Concerning the two DL models, they have been used, with different variations, in previous studies on the automatic detection of software vulnerabilities: CNN \cite{RUSSELL, FWU2017, Li20182, HARER2018} and RNN \cite{RUSSELL, Li2018, Li20182, LIN2019}. We also exploit as baseline for our experiments an automated machine learning (AutoML) approach, which is a solution to build DL systems without human intervention and not relying on human expertise. AutoML has been used in Natural Language Processing (NLP) and it is provided by Google Cloud Platform (GCP). AutoML eases the hyper-parameter tuning and feature selection using Neural Architecture Search (NAS) and transfer learning \cite{he2020automl,wong2018,ZOPH2017}. 

Since we are experimenting with different source code representations, datasets and type of classification (binary and multi-class), we run a total of 30 different experiments for the AutoML-based baseline approach and 90 experiments for the deep/shallow approaches. The explanation and details of such a variety of experiments is provided later on in this section.


The \emph{context} of the study is represented by three datasets of C/C++ code reporting software vulnerabilities at the function granularity level, for a total of 1,841,323 functions, of which 390,558 are vulnerable ones. Our study addresses the following research question (RQ):

\begin{quote}
	\emph{\rqone}
\end{quote}

We answer this RQ in two steps. First, we create binary classifiers able to discriminate between vulnerable and non-vulnerable functions, without reporting the specific type of vulnerability affecting the code. This scenario is relevant for practitioners/researchers who are only interested in identifying potentially vulnerable code for inspection/investigation. Second, we experiment the same models in the more challenging scenario of classifying functions as \emph{clean} (\ie do not affected by any vulnerability) or as affected by specific vulnerabilities.

\begin{table}[t]
	\caption{Number of functions in each subject dataset.}
	\label{tab:datasets}
	\centering
	\begin{tabular}{lrrr}
		\toprule
		& GH-DS & J-DS & R-DS\\
		\midrule
		Vulnerable & 315,777 & 28,446 & 46,335\\
		Non-vulnerable & 315,777 & 62,475 &  1,072,513 \\\bottomrule
	\end{tabular}
	\vspace{-0.4cm}
\end{table}

\subsection{Data Collection}
We relied on three datasets (\tabref{tab:datasets}) composed by C/C++ functions and information about the vulnerabilities affecting them. The datasets are described in the following.

\textbf{GitHub Archive Dataset (GH-DS).} We built GH-DS starting from GitHub Archive \cite{GHA}, a dataset containing every public GitHub event (\eg commits, opening of pull requests) from March 2011. 

Since the events generated before 2015 were stored using a deprecated data format, we focused on public events going from January 2015 up to November 2018, when we started the building of GH-DS. We mined from this list of events all vulnerability-fixing commits accompanied by a message containing the exact (i) name of a vulnerability as reported in the CWE dictionary (\eg Use After Free) and/or (ii) id of a CWE vulnerability in the format CWE-416. We assume that in these commits developers are fixing the vulnerability described in the commit message. However, as we are aware of the fact that commit messages might imprecisely identify bug-fixing commits \cite{AntoniolAPKG08,HerzigJZ13} and, as a consequence, vulnerability-fixing commits, two authors independently analyzed a statistically significant sample (95\% confidence level $\pm$5\% confidence interval, for a total size of 384) of identified commits to check whether they were actually vulnerability fixes. After solving 45 cases of disagreement, they concluded that 90.3\% of the identified vulnerability-fixing commits were true positives (additional details  in our replication package \cite{replication}). For each vulnerability-fixing commit, we extracted the source code before and after the fix using the GitHub Compare API \cite{github-compare}. This allowed us to collect the vulnerable (pre-commit) and the non-vulnerable (post-commit) code.

We discarded all commits not modifying at least one C/C++ file and those not available anymore on GitHub. We also discarded commits modifying more than one file, to be sure about the ``context'' of the vulnerability fix. This reduces the likelihood of including tangled commits \cite{Herzig:msr2013} comprising a vulnerability fix as well as other changes (\eg some refactoring operations). Finally, we used GumTreeDiff \cite{GUMTREE} to identify the list of edit actions performed between the vulnerable and the fixed C/C++ files in each commit. In this way, we identified the functions modified to fix the vulnerability, and we built pairs of functions ($f_v$, $f_nv$) where $f_v$ represents the vulnerable version of function $f$ (pre-commit) and $f_nv$ represents the fixed (non-vulnerable) version of $f$ (post-commit). 

The final GH-DS dataset is composed of $\sim$315k pairs of vulnerable and non-vulnerable C/C++ functions related to the fixing of 70 different types of vulnerabilities across $\sim$29k GitHub projects. We release this dataset to the research community \cite{replication}. While we use it to experiment vulnerability detection techniques, GH-DS can also be used for the development of approaches to automatically fix vulnerabilities. Indeed, having available the version of the same function before and after the fix of a vulnerability could allow the usage of deep learning techniques such as Neural Machine Translation (NMT) to learn the code changes needed to ``translate'' a vulnerable function into its non-vulnerable version.

\textbf{STATE IV Juliet Test Suite Dataset (J-DS) \cite{JULIET}.} This existing dataset has been used in previous studies on automated vulnerability detection \cite{RUSSELL} and includes synthetic code examples (\ie the vulnerabilities have been manually injected) of 118 types of CWE vulnerabilities. For each example, both the vulnerable and the non-vulnerable versions of the code are available. We did not use the dataset as provided but performed a few cleaning steps aimed at excluding (i) vulnerabilities that are not at function-level (\eg \texttt{CWE401\_Memory\_Leak\_\_destructor\_01\_bad.cpp}) and (ii) single vulnerability instances that are spread across more than one function, since in this study we want to investigate the potential of vulnerability detection techniques in the simplest 1-to-1 scenario: One function is clean or it is affected by a single type of vulnerability \textbf{and} a vulnerability instance does only affect a single function. 

\textbf{Russell \etal Dataset (R-DS).} This dataset is available as part of the work by Russell \etal~\cite{RUSSELL}, and was created by using functions from the Debian Linux distribution and GitHub public repositories. The authors collected $\sim$1.3M functions and used three static analysis tools (\ie Clang \cite{CLANG}, Cppcheck \cite{CPPCHECK}, and Flawfinder \cite{FLAWF}) to tag them with a label identifying the type of vulnerability affecting the code (if any). The authors considered six types of labels. Four are related to the most popular types of vulnerabilities they found, \ie CWE-119 (Improper Restriction of Operations within the Bounds of a Memory Buffer), CWE-120 (Buffer Overflow), CWE-469 (Use of Pointer Subtraction to Determine Size), and CWE-476 (NULL Pointer Dereference). 

One is a macro category identifying functions affected by other, less popular, vulnerabilities (\ie CWE-others). Finally, the last label identifies the clean functions. 2\% of the functions were tagged with multiple vulnerabilities; 4\% of the functions were tagged with one specific type of vulnerability; the rest of the functions (94\%) were classified as not vulnerable, which makes it a highly unbalanced dataset. We excluded also from this dataset the functions tagged with multiple vulnerabilities for the reasons previously explained. 

\subsection{Code Representation}
From each dataset (GH-DS, J-DS, R-DS) we extracted two sets of tuples. The first one, in the form $\langle function$\_ $code, is$\_$vulnerable \rangle$, aims at experimenting the models in the scenario in which we want to identify vulnerable functions, but we are not interested to the specific type of vulnerability. In the second, the tuples are instead in the form $\langle function$\_ $code, vulnerability$\_$type \rangle$, to experiment the models in the scenario in which we want to classify the vulnerability type exhibited by a given function. We use the $non$\_ $vulnerable$ label to identify functions not affected by any vulnerability. 

We built five versions of each set of tuples (\ie binary and multi-class) representing the code in different ways, to study how the code representation affects the performance of the models. The generation of the representations was supported by a custom C/C++ lexer we wrote to extract the needed information (\ie the syntactic and semantic meaning of the tokens in each function). Using the raw source code as input for machine learning techniques is not an option especially due to the huge vocabulary of terms used in the identifiers and literals. Such a large vocabulary would hinder the goal of learning patterns in the data that characterize vulnerable functions. For this reason, we abstract the code and generate expressive yet vocabulary-limited representations, as done in previous applications of machine learning on code \cite{Allamanis:MSR13,White2015,Tufano:icse2019}.

\begin{figure}[t]
	\centering
	\includegraphics[width=\linewidth]{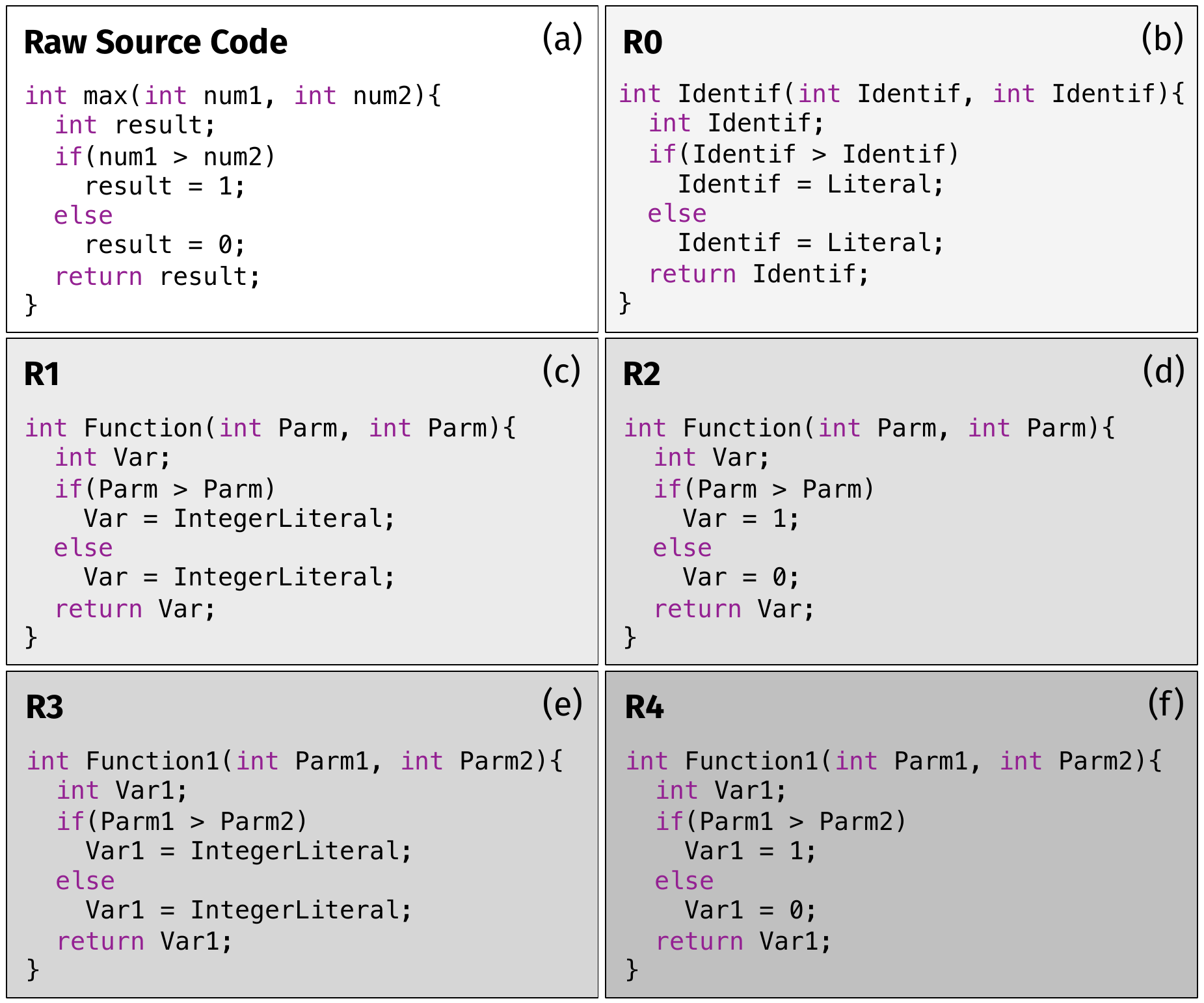}
	\caption{Code representations used in our study (b) to (f). Subfigure (a) shows a sample raw source code provided as input.}	
	\label{fig:representations}
\end{figure}

In the following, we describe each representation using as reference the source code snippet presented in \figref{fig:representations}-(a).

\textbf{R0: Baseline.} This is the most basic representation, since it provides minimum information in terms of code context. We include in R0 the C/C++ reserved keywords and punctuation symbols, while identifiers and literals are represented with the tokens \texttt{Identifier} and \texttt{Literal}, respectively.  \figref{fig:representations}-(b) depicts the function in \figref{fig:representations}-(a) using R0 as representation.

\textbf{R1: Augmented R0.} As in R0, C/C++ keywords and punctuation symbols are kept, but identifiers are tokenized according to their type (\eg \texttt{Function}, \texttt{Parameter}, \texttt{Variable}). Also, for tokens related to literals, we explicitly label the ones related to strings as \texttt{StringLiteral} and those related to integer values as \texttt{IntegerLiteral}. 

All the others are just kept as \texttt{Literal}. The rationale for this choice is that string and integer literals are involved in security vulnerabilities because, on the one side, strings can be used to represent permissions, access rules, and file paths; on the other side, integers are often used for buffer sizes, iterators indexes/limits, and offsets that play a role in different types of vulnerabilities. 
\figref{fig:representations}-(c) depicts the function in \figref{fig:representations}-(a) using R1 as representation.

\textbf{R2: R1 + Integer values.} This representation is almost the same as R1, but all one digit numeric literals are kept in this case.  \figref{fig:representations}-(d) shows the R2 representation of the function used as running example. The rationale behind this representation is that integers are responsible for many types of vulnerabilities (\eg CWE-119, CWE-125, CWE-129, CWE-369, CWE-469, CWE-665, CWE-682, CWE-839) \cite{Seacord}. 

\textbf{R3: R1 + Identifiers Numbering.} Also R3 is built on top of R1. In this case, instead of listing all identifiers with their type, we keep track of the identifiers by numbering them, as done by Tufano \etal \cite{Tufano:icse2019}. Assume that two variables are present in a function, and these variables are used several times inside the function body. With the R1 representation, all usages of these variables will be replaced with the \texttt{Var} token. In R3, all the instances of the first variable will be replaced with the \texttt{Var1} token, and all those of the second variable will use the \texttt{Var2} token (see \figref{fig:representations}-(e)). This representation aims at ``informing'' the classification model about the fact that the same identifier is used in different parts of the function body which could allow for identifying  data flow dependencies.

\textbf{R4: R2 + Identifiers Numbering.} This is the representation providing more contextual information: We start from R2 and integrate the identifiers' numbering previously explained for R3. \figref{fig:representations}-(f) depicts this representation using as reference the example in \figref{fig:representations}-(a).

\input{secs/sb_cleaning_classifiers_analysis}

%% file: secs/sb_cleaning_classifiers_analysis.tex
\subsection{Data Cleaning}
Before using the three datasets to train and evaluate the experimented models, we performed a transformation and cleaning process on each of them to (i) make the data treatable with DL/shallow models, and (ii) avoid possible pitfalls that are common in studies of machine learning on code (\eg duplicated functions due to forked projects \cite{Allamanis:MSR13, Allamanis:18}). 

First, for each dataset we created different versions using the five code representations shown in \figref{fig:representations}. This gave us five different datasets of function representations. Second, we addressed  conflicting representation (\ie two samples with same code representation and different labels) and duplicates. In case of conflicting representations, all instances were removed. As for the duplicates, we removed all duplicates having the same raw source code representation and the same label (\ie type of vulnerability affecting them, if any), keeping only the first occurrence. This means that it is possible to have in our datasets two snippets having the same abstract representation, but not the same raw source code. Such a design choice is justified by the fact that the translation from raw source code to abstract representation is part of the classification pipelines used in  ML implementations, and it is performed after the removal of duplicates.

Duplicates and conflicting instances can be different in the same dataset when using different representations. For example, two functions can be different using the more expressive R4 representation and equal when using the basic R0 representation. In this case, the duplicate will be removed from the dataset only when using the R0 representation. \tabref{tab:duplicatedData} summarizes the final number of functions in each dataset after this process. It is important to mention that after this process GH-DS is no longer balanced.


\begin{table}[t]
	\caption{Number of functions in each dataset for each representation after handling duplicates and conflicting representations.}
	\label{tab:duplicatedData}
	\centering
	\begin{tabular}{lrrr}
		\toprule
		 & GH-DS & J-DS & R-DS\\
		\midrule
		R0  & 164,919 & 56,073 & 1,117,137\\
		R1  & 166,187 & 56,075 & 1,117,765\\
		R2  & 167,280 & 56,083 & 1,117,925\\
		R3  & 172,421 & 56,075 & 1,118,349\\
		R4  & 173,498 & 56,083 & 1,118,488\\
\bottomrule
	\end{tabular}
\end{table}

Third, for each dataset, we computed the distribution of number of tokens for all function representations it contains. Then, for each dataset representation, we excluded functions that are shorter than the first quartile (\eg for R1 J-DS 59 tokens) or longer than the third quartile (\eg for R1 J-DS 115 tokens). This has been done to exclude very short and very long functions from our dataset since the former provide very little information to characterize the presence of a vulnerability and the second are computationally expensive to process with DL models. Removing  upfront very long functions, also prevents from truncating sentences (in our case, functions) to a fixed maximum length during the training, which would provide incomplete inputs to the model, something suboptimal when working with source code. After this step, the data cleaning process for binary classification ends, and the final number of functions (F) for this type of classification is reported in \tabref{tab:cleanedDataB}, which reports also the percentage of duplicates in the abstract representation. 

Remember that duplicates can be in the abstract representation but not in the raw source code. The maximum times a single sample repeats is 18 for J-DS and 77 for GH-DS.


\begin{table}[t]
	\caption{Number of functions in each dataset for each representation after data cleaning. Total number of functions (F), percentage of duplicates (D)}
	\label{tab:cleanedDataB}
	\centering
	\scalebox{0.95}{
	\begin{tabular}{l|rr|rr|rr}
		\toprule
		& \multicolumn{2}{c|}{GH-DS} &  \multicolumn{2}{c|}{J-DS}  &  \multicolumn{2}{c}{R-DS} \\
		&F& D &F& D &F& D \\
		\midrule
		R0  & 70,618 &26.61\%& 28,572 &36.18\%& 563,646&0\%\\
		R1  & 71,757 &26.52\%& 28,572 &34.97\%& 563,752&0\%\\
		R2  & 72,214&26.42\% & 28,572 &34.93\%& 563,817&0\%\\
		R3  & 79,216 &28.26\%& 28,572 &34.63\%& 563,923&0\%\\
		R4  & 79,545 &28.19\%& 28,572 &34.58\%& 563,970&0\%\\
		\bottomrule
	\end{tabular}
}
\end{table}


Finally, to have enough data points to train the multi-class models in charge of identifying the specific type of vulnerability affecting a given function, we kept in each dataset only the five most represented vulnerabilities and the samples belonging to the ``non-vulnerable'' class (see \tabref{tab:topFive}). 

Note that this was only needed for GH-DS and J-DS, since R-DS does already focus on four possible categories of vulnerabilities. In R-DS one of the possible categories (CWE-119), after the cleaning process, had only five samples and we decided to merge it with another category (CWE-120), since representing similar vulnerabilities. \tabref{tab:cleanedDataM} summarizes the final number of functions for multi-class classification.

\begin{table}[t]
	\caption{Top five vulnerabilities in each dataset. *Russell has only four vulnerabilities, and two of them were merged}
	\label{tab:topFive}
	\centering
		\scalebox{0.95}{
	\begin{tabular}{p{0.06 \textwidth}p{0.37  \textwidth}}
		\toprule
		 Dataset& Top Vulnerabilities \\
		\midrule
		GH-DS  & Deadlock, Race Condition, Null Pointer Dereference, Buffer Overflow, Dead Code\\
		J-DS   & CWE-401, CWE-762, CWE-590, CWE-122, CWE-121 \\
		R-DS*   & CWE-other, CWE-120, CWE-476\\
\bottomrule
	\end{tabular}}
\end{table}


\begin{table}[t]
	\caption{Number of functions in each dataset for each representation after data cleaning for multi-class classification. Total number of functions (F), percentage of duplicates (D).}
	\label{tab:cleanedDataM}
	\centering
	\scalebox{0.95}{
	\begin{tabular}{l|rr|rr|rr}
		\toprule
		& \multicolumn{2}{c|}{GH-DS} &  \multicolumn{2}{c|}{J-DS}  &  \multicolumn{2}{c}{R-DS} \\
		&F& D&F& D &F& D \\
		\midrule
		R0  & 57,475 & 33.19\%& 11,445 &35.98\%& 563,646&0\%\\
		R1  & 57,947 &32.76\%& 11,445 &34.95\%& 563,752&0\%\\
		R2  & 58,390 &32.70\%& 11,445 &34.38\%& 563,817&0\%\\
		R3  & 60,305 &31.00\%& 11,445 &34.49\%& 563,923&0\%\\
		R4  & 60,768 &31.00\%& 11,445 &34.38\%& 563,970&0\%\\
		\bottomrule
	\end{tabular}
}

\end{table}

At the end of this process we obtained 30 different datasets. The first 15 datasets (five representations times three datasets) correspond to the binary classification, whereby the labels can only take two possible values (vulnerable or non-vulnerable). The other 15 datasets correspond to the multi-class classification and their labels represent a specific type of vulnerability or a non-vulnerable instance. 

\subsection{Classifiers}
\label{subsec:classifiers}

Given the four approaches (\ie GCP-AutoML, RF, CNN \& RNN), five representations, three datasets, and two types of classification (binary and multi-class) used in our study, we built a total of 120 different models ($4*5*3*2$).


As previously mentioned, we started by training our shallow learning model (RF), the baseline (GCP-AutoML) and the two aforementioned deep learning models (CNN and RNN). Regarding GCP-AutoML, the training  process is the same for both binary and multi-class classification and it consists of only two steps: (i) we upload a given dataset featuring a function representation (\eg R0 for J-DS) with its corresponding label (\eg vulnerable); and (ii) we trained a model with the GCP implementation of AutoML which uses NAS \cite{le_zoph_2017,ZOPH2017}. NAS uses different optimization algorithms to find an optimal neural network architecture without requiring human expertise \cite{ZOPH2017}. Thus, for the GCP-AutoML representing our baseline we did not need to define the type of model, the search space for the hyper-parameters, nor to process the input from text to sequence tokens. This is all automatically handled.

Concerning the RF, for binary and multi-class classification the model is the same, with the only difference being the labels in the training/test set. 

We used a balanced class weighting to contrast imbalanced classes. For this model the input data is a Bag of words (BOW) of the specific representation and the output is a label classifying the instance. We used the Scikit learn \cite{sklearn_api} implementation and the hyper-parameters we tuned are reported in \tabref{tab:HyperParameters}.

\begin{table}[t]
	\caption{Hyper-parameters that were tuned for each type of model.}
	\label{tab:HyperParameters}
	\centering
	\begin{tabular}{p{0.06 \textwidth}p{0.37  \textwidth}}
		\toprule
		Model & Hyper-parameters tuned\\
		\midrule
		RF  & Number of estimators, Number of features to consider at every split, Maximum  number of levels in tree, Function used to evaluate the quality of a split\\\\
		CNN  & Dimension of the dense embedding, number of Convolution layers, Number of output filters, Activation function for Convolution layer, Initializer for the kernel weights matrix, Stride Length of the convolution, Number of dense layers, Drop out rate, Activation function for each dense layer, Batch size\\\\
		RNN  & Dimension of the dense embedding, Number of LSTM layers, Activation function  for recurrent layer,Number of units of the recurrent layer, Initializer for the kernel weights matrix, Regularizing factor for the kernel weights matrix, drop our rate, Number of dense layers, Dense Layers Activation function, Batch size, Optimizer's learning rate
		\\
		\bottomrule
	\end{tabular}
\end{table}

For RNN and CNN models, the input layer takes as input a token sequence extracted from the specific code representation and converts it into a fixed $k$-dimensional  embedding representation. The $k$ value is one of the hyper-parameters to tune. The output layer is a one-hot encoded vector that corresponds to a particular label. For the binary case the output layer has a sigmoid activation function and one neuron. 

For the multi-class scenario it uses a softmax activation function and the number of neurons is equivalent to the number of prediction classes (\eg for R-DS the output layer has 4 neurons, but for J-DS and GH-DS it uses six neurons). The hidden layers between the input and output layers differ between these two models.

The RNN has LSTM layers after the embedding layer. The hyper-parameters we tune for this layer are: the number of units and the initialization of the kernels and the activation function. After this layer, the RNN uses a drop out layer to prevent overfitting followed by dense layers.

For the CNN, after the embedding layer, we have  1D dimensional convolutional layers, since we are working with one dimensional data (text), as opposed to other contexts in which more dimensions are needed (\eg to manipulate images). The hyper-parameters that we tune for this layer are:  the number of filters, the kernel size and the activation function. After this layer, the CNN model includes a global max pooling layer, a drop out layer (to reduce the chance of overfitting) with a rate that can be tuned, and one or two dense layers depending on the hyper-parameter optimization process.

For both CNN and RNN, the learning rate and the batch size were also hyper-parameters to tune. We used the Adam \cite{kingma2017adam} optimizer. 
For both models we used Keras \cite{chollet2015keras}. The set of tuned hyper-parameters is reported in \tabref{tab:HyperParameters}.

\input{secs/analysis}

%% file: secs/analysis.tex

\subsection{Analysis Method}
We split each of the 30 datasets into training (60\%), validation (20\%), and test (20\%) set using stratified random sampling. This strategy ensures that the labels frequency distribution is approximately equal between the sets and similar to the original distribution. For GCP-AutoML, we uploaded a \textit{.csv} file with the following information for each sample: type of set  (\ie train, validation, test) it belongs to, its abstract representation, and its label (\eg vulnerable).  The datasets used for GCP-AutoML are the same used in our models, with the only difference that GCP-AutoML requires to remove the duplicates even at abstract representations.

Since we had two different families of approaches (\ie shallow and deep learning) which expect a different type of input (\ie BOW \emph{vs} sequence of tokens), it was necessary to add a lightweight preprocessing to the data. For the RF, we generate, starting from the code representation, a TF-IDF BOW to weight the importance of each token in the BOW. 

For the CNN and RNN models, it was necessary to convert each text representation into a sequence of tokens of the same length. All the instances in each dataset that were shorter than the upper limit were post padded: We pad sequences with a special token until it reaches a fixed length.

Training and validation sets were used to build the models and to tune the hyper-parameters of each combination of classifier and code representation, while the test set was used for assessing the performance of the models. When tuning CNN, RNN and RF we used a Bayesian-search \cite{BAYESIAN2012, BY2}. This type of optimization performs a directed search based on the surrogate function and an acquisition function. A surrogate function is a model that is used for approximating the objective function, while the acquisition function directs sampling to combinations of parameters where an improvement over the best observation is likely. For this process we used the Optuna  \cite{optuna_2019} Python library.

In the binary classification scenario, the three models  (\ie RF, CNN, RNN) were trained with cross entropy loss, while for multi-class classification we adopted the categorical cross entropy loss. In the tuning process we optimized accuracy for binary classification and  categorical-accuracy for multi-class.

Once each model was tuned, its performance has been evaluated on the test set. We computed \textbf{\textit{Accuracy}} and Matthews Correlation Coefficient (MCC) \cite{MATTHEWS1975, Baldi2000, GORODKIN2004367} for binary and multi-class scenarios. Accuracy evaluates the overall performance of the models while MCC was used to evaluate the performance with a metric that is less influenced by imbalanced data \cite{boughorbel_optimal_2017, he_learning_2009,jeni_facing_2013}. Further, we also compute Precision, Recall and F1 for the binary classification scenario, having vulnerable functions as the positive class. For the multi-class scenario we compute macro averages for Precision, Recall and F1. For the GCP-AutoML model we calculate the aforementioned metrics based on the confusion matrix that GCP returns for the test set.

%% file: secs/results.tex
\section{Results}
\label{sec:results}
We start by discussing the results achieved by the binary classifiers, having the goal of discriminating between vulnerable and non-vulnerable functions. 

\begin{figure*}
	\centering
	\includegraphics[angle=0,width=0.98\textwidth]{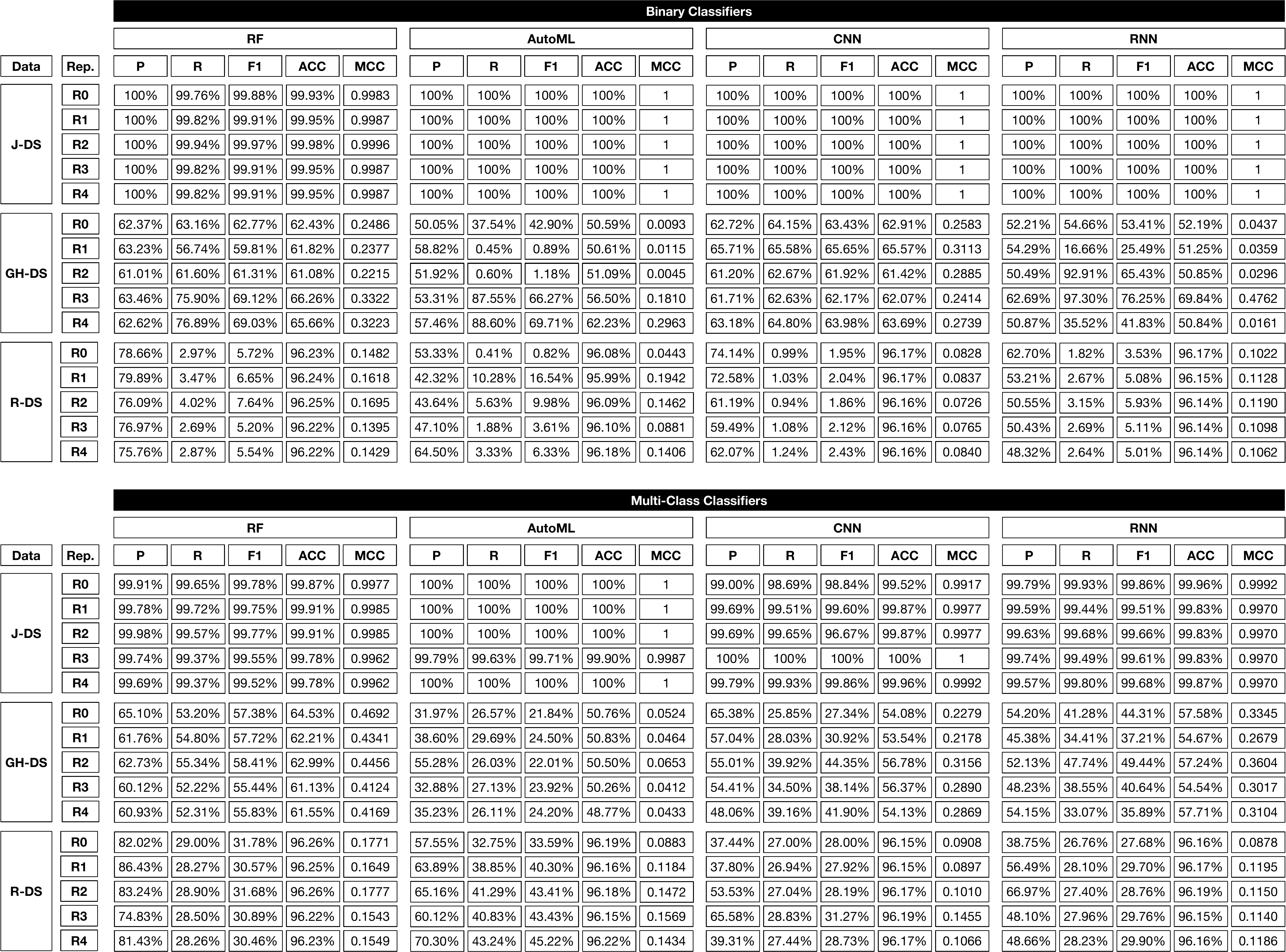}
	\caption{Top: Binary Classification metrics: Precision (P), Recall (R), F1, Accuracy (ACC), and MCC for each approach and dataset-representation, using vulnerable as the positive class; Bottom: Mutli-class Classification metrics: Macro Precision (P), Macro Recall (R), Macro F1, Accuracy (ACC), and MCC for each approach and dataset-representation}	
	\label{representations}
\end{figure*}

Then, we present the findings related to the multi-class classifiers, which classify functions as non-vulnerable or as affected by a specific type of vulnerability. The results achieved by the models for different representations of the code should not be directly compared, since small differences exist in these datasets due to the removal of duplicates. However, within the same representation and dataset, the performance of the different models can be compared. AutoML, due to its peculiarities, only serves as baseline to assess the performance a model only requiring a minimum human intervention is capable of achieving for this task. The results achieved by the experimented models can be found in \figref{representations}. 

\subsection{Binary Classification}
In binary classification, the models can be used to flag functions likely to be vulnerable, for a subsequent code review performed by developers. We discuss three possible scenarios: (\RN{1}) Recall is favored over precision, meaning that the cost of reviewing the vulnerable functions flagged by the models is considered low by the developers (\ie a high number of functions to check  flagged as potentially vulnerable is acceptable) while the potential cost of a missed vulnerable function is high (\ie it is not acceptable to release the code with a vulnerable function); (\RN{2}) Precision is favored over recall, meaning that the manual inspection cost is considered high and the risk of missing a vulnerability is considered limited; (\RN{3}) Balancing precision and recall is desirable, with metrics such as F1, MCC or Accuracy that should be maximized. 

Before presenting the best models for each of these scenarios, it is important to discuss the results achieved for the J-DS dataset (top part of \figref{representations}): All models exhibit extremely high performance, with the three DL-based being perfect predictors (\ie MCC=1) and the RF being very close (MCC$>$0.99). 

These results, especially when compared with the ones achieved on the other two datasets, clearly highlight potential issues with the synthetic origin of this dataset, that includes artificially created instances of vulnerable functions. It is worth noting that this dataset has been used in previous works on vulnerability detection \cite{RUSSELL, Li2018}. From a manual inspection of its instances, we found that almost all non-vulnerable samples are characterized by the presence of the $static$ token, that makes the classification task trivial for the models. This poses doubt on the usefulness of this dataset for assessing the performance of vulnerability detection approaches.

Moving to the GH-DS dataset that, as explained in \secref{sec:design}, was built by using real instances of vulnerable functions mined from GitHub; the results achieved by the models with GH-DS, as expected, substantially drop. Overall, the best models are the RF and the CNN that achieve very similar performance. Considering the higher training cost of the CNN classifier, a RF could be preferred in many scenarios. 

Also, while RF and CNN are undisputedly the best model in the scenario \RN{2} (maximize precision) and \RN{3} (maximize overall accuracy)\footnote{The only exception to this trend is the RNN with the R3 representation.}, this is not always the case for \RN{1} (maximize recall). 

Indeed, here the RNN model and the AutoML model manage to achieve better results for specific code representations..

Concerning the third dataset (\ie R-DS) we also obtained more ``realistic'' results as compared to J-DS, with all models struggling to obtain high values for recall. Indeed, independently from the used model and code representation, the recall is usually lower than 5\%. In this scenario, the RF seems to be the best solution. Indeed, while also for it the recall is extremely low, at least the shallow model is the one able to achieve the best precision values.

\textbf{Concluding remarks.} 
The results achieved by our baseline (AutoML), are inline or lower as compared to the other three experimented approaches. This serves as a sort of sanity check for our implementations, confirming how challenging the tackled problem is (\ie automated vulnerability detection) even in the simpler scenario (\ie binary classification). 


Despite RF being representative of a shallow learning model, it achieved performance comparable or, in some cases, even better (see the R-DS) than the deep learning models. 

The result achieved on J-DS can instead be basically ignored, if not for the message they give to the research community about the usage of this dataset in the assessment of vulnerability detection tools. 



\begin{figure*}
	\centering
	\includegraphics[angle=0,width=0.8\textwidth]{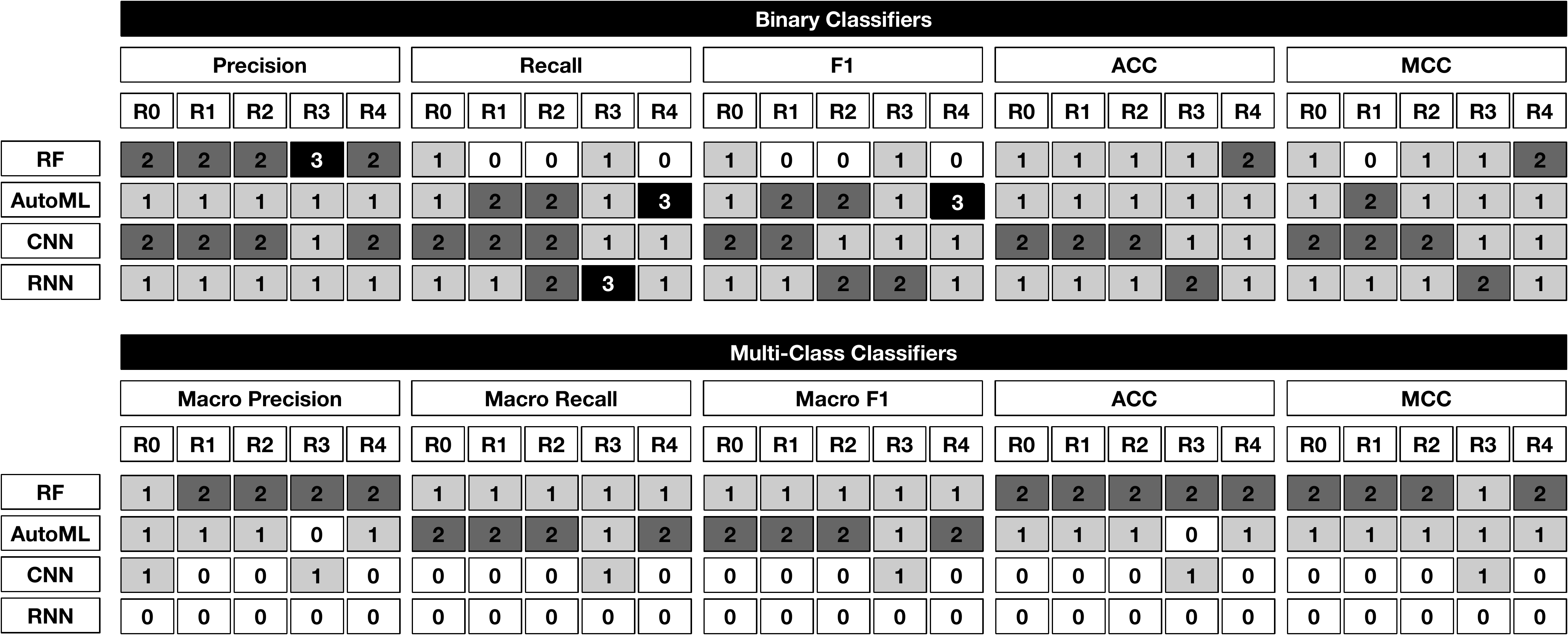}
	\caption{Heatmap conveying the amount of times a model obtained the greatest score given a metric. Top: Binary Classification metrics; Bottom: Mutli-class Classification metrics.
	}	
	\label{heatmap}
\end{figure*}

To summarize our findings, we also present in \figref{heatmap} a graphical representation conveying the number of times a model obtained the greatest score for each metric (\eg Precision) given a representation. \figref{heatmap} shows that (i) the RF is the better performing in terms of precision; and (ii) CNN is the most ``balanced'' across all considered metrics.



\subsection{Multi-class Classification}

\input{secs/resultsMulti}


%% file: secs/resultsMulti.tex

In the multi-class classification (bottom part of \figref{representations}),  we have two main scenarios: (\RN{1}) The first looks at the overall performance of the models (\ie accuracy), without considering the imbalanced nature of the subject datasets (\ie a vulnerability type might be more frequent than another); (\RN{2}) In the second, by analyzing the MCC we can evaluate how well a model scores considering the unbalanced datasets we deal with. 

The first scenario is useful to give higher weight to the majority class (non-vulnerable functions) when computing the performance of the model; the second looks for a model that is able to provide good performance across all classes. As already observed for the binary classification problem, the results achieved by all models on the J-DS dataset are extremely high, confirming the unchallenging classification task of this dataset. 


Concerning GH-DS, RF is consistently the best model, achieving the greatest values for all the scores. RF also obtained very balanced macro-precision and macro-recall values, that differ less than 7\%, being a model that does not sacrifice precision for recall or \emph{vice versa}. Furthermore, RF performs well both when considering accuracy ($>$60\%) and MCC ($>$0.4) as performance proxy.

Finally, also for the R-DS dataset, the RF was overall the best model in terms of both MCC and Accuracy, with the exception of AutoML with the R3 representation. However, the difference in performance in favor of AutoML is relatively small, while its higher training cost is substantial as compared to the RF. It is also worth noticing that all models on this dataset (R-DS) achieve a high accuracy in all the cases ($>$96\%). This is due to the fact that this dataset is highly imbalanced (see \tabref{tab:datasets}) and a constant classifier always predicting the functions as ``non-vulnerable'' would achieve, on this dataset a $\sim$95\% accuracy. Thus the results in terms of MCC are more representative for this dataset.

\textbf{Concluding remarks.} 
Shallow learning using RF performed well across all different datasets, being the best model in the multi-class scenario. It is important to highlight that this finding holds for the specific DL-models, settings, datasets, and code representations we used. Our study does not claim the superiority of shallow over deep learning, but rather the fact that shallow learning baselines such as the RF, properly tuned, should always be considered as a valid baseline to compare with for vulnerability detection. 

Also in this case, we conclude our results discussion presenting the bottom part of the heatmap in \figref{heatmap}. 

The RF confirms its superiority both in terms of accuracy and MCC, with the RNN  not being the best model for any of the investigated representations and metrics.

%% file: secs/threats.tex
\vspace{0.2cm}
\section{Threats to Validity}

	Threats to \textbf{construct validity} concern the relation between the theory and the observation. When building GH-DS we identified vulnerability-fixing commits based on commit messages, assuming that the developer is fixing the vulnerability described in the commit message. However, it is known that parsing commit messages might imprecisely identify bug-fixing commits \cite{AntoniolAPKG08,HerzigJZ13} and, for similar reasons, vulnerability-fixing commits. To mitigate subjectivity bias, two authors independently analyzed a statistically significant sample of 384 commits identified by our heuristics as vulnerability-fixing commits. After solving conflicts, they found that 347 of them represented true positives (\ie correctly identified commits). 
	
		As previously explained, the results achieved by the models for different representations of the code should not be directly compared, since small differences exist in these datasets due to the removal of duplicates. However, within the same representation and dataset, the performance of the different models can be compared, and we did that by looking at a wide set of metrics assessing different aspects of the classification. 

	Threats to \textbf{internal validity} concern external factors we did not consider that could affect the variables and the relations being investigated. We applied stratified random sampling when splitting the dataset into training (60\%), validation (20\%), and test (20\%) set to ensure that the labels frequency distribution was similar among the sets and to the original distribution.
	
	An important factor that influences the models' performance is calibration of hyper-parameters, which has been performed as detailed in \secref{sec:design}. We are aware that additional tuning, possibly performed by using a different search strategy, could produce different and even better performances, especially for the DL models. 
	
	It is also important to highlight that a substantial difference exists between the tuning and training process performed for the three models subject of our study (\ie RF, RNN, and CNN) as compared to the AutoML baseline. For the former, we defined the search space for hyper-parameters tuning and we performed the preprocessing needed to transform our original abstract representations into a tokens sequence or a BOW representation. 
	
	As for AutoML, we just uploaded the original abstract representations to the GCP, with the latter taking care of modifying the input if needed, deciding also the hyper-parameters to be tuned. While AutoML reduces the human interaction needed in the process of finding a good model, it does not allow to get insights about the model's architecture and, for this reason, we had to treat it as a black-box. 

Threats to \textbf{conclusion validity} concern the relationship between treatment and outcome. We adopted metrics allowing to obtain a clear overview of the performance provided by the models in the experimented scenarios. For example, in the case of multi-class classifiers, taking only the Accuracy into account does not provide a picture of how well the model works across all classes involved in the classification (\eg the different types of vulnerabilities). Furthermore, in multi-class classification it is important to document the type of average adopted to avoid a misunderstanding of the models performance: macro-average gives the same importance to each class being insensitive to class imbalance, while micro-average maximizes hits.

	Our results concerning J-DS and R-DS cannot be directly compared with those reported in previous works sharing some of the datasets we used \cite{Li2018, RUSSELL}. This is due to several reasons. First, the models are not being trained with the same amount of samples (\ie number of functions). Second, the code granularity varies: Li \etal \cite{Li2018} use \textit{code gadgets}, \ie a number of (not necessarily consecutive) lines of code that are semantically related to each other, as opposed to our function-level granularity. Third, the classification approaches differ from ours: While we studied binary and multi-class classification, Li \etal \cite{Li2018} train multiple binary classifiers, one for each type of vulnerability they studied, instead of a multi-class classifier. Finally, the adopted performance metrics differ among the discussed studies. For example the type of multi-class averaging adopted is not documented in those studies. Thus, our study should not be considered as a replication of previous work, but as complementary to past studies.

	With respect to duplicate handling, we are aware and documented the presence of duplicates in our datasets due to the  translation procedure we adopted from raw source code to the different abstract representations. Still, we removed duplicates at raw source code level, but intend to investigate in future how the performance of the approaches change when removing duplicated abstracted instances. In relation to duplicates handling for the AutoML baseline, since this utility is a black-box from the user point of view and it expects a dataset without duplicates we had to remove the duplicated abstracted instances to avoid receiving processing errors.

	Threats to \textbf{external validity} concern the generalizability of our findings. We conducted our study with several variables involved, namely four approaches (\ie GCP-AutoML, RF, CNN \& RNN), five code representations, three datasets, and two types of classification---binary and multi-class). This required building a total of 120 models ($4*5*3*2$). Also, our datasets are considerably different, including both synthetic instances and real-world vulnerable instances mined from open source projects. Overall, the three datasets account for $\sim$1.8M functions. However, all datasets focus on C/C++ code and report software vulnerabilities at function level. Therefore, the achieved results may not be valid for other languages and/or different granularity levels.

%% file: secs/conclusion.tex

\section{Conclusions}
We presented a large-scale empirical study aimed at analyzing the effectiveness of deep and shallow learning techniques for detecting software vulnerabilities in source code. Our study is based on the analysis of three datasets of C/C++ code reporting software vulnerabilities at the function granularity level, accounting for a total of $\sim$1.8M functions, of which $\sim$400k are vulnerable. 

Our results show that there is large room for improvement in the field of automated vulnerability detection, independently from the usage of shallow or deep models. Indeed, our results show that shallow models such as the Random Forest can achieve competitive  performance for the specific task we investigated.

Our future work will be mostly driven by the findings discussed in \secref{sec:results}. First, we are planning to expand our study by building datasets featuring other programming languages in order to obtain more generalized conclusions. Second, while we adopted a simple code representation based on streams of tokens for the DL-based systems, more complex representations tailored for source code have been proposed recently (see \eg \cite{code2vec}); we plan to investigate the possible benefits brought by these representations in our models. We will also experiment with other models such as ensembles, BLSTM, GRU, and transformers \cite{vaswani2017transformer}. Another avenue for future work is the analysis of the  impact  of different combinations of  code granularities and representations in the detection models.  Finally, we will extend the study to include the multi-label case, \ie code units annotated with more than one vulnerability.

%% file: ack.tex

\section*{Acknowledgment} \label{section:ack}

Mazuera-Rozo and Bavota gratefully acknowledge the financial support of the Swiss National Science Foundation for the CCQR project (SNF Project No. 175513). Linares-Vásquez is partially funded by a Google Latin America Research Award (LARA) 2018-2021, and the Uniandes Vice-rectory for Research under the CI-120 call for grants.